\documentstyle[11pt,equations]{article}
\begin{document}
\newcommand{\tcr}{T_{cr}}
\newcommand{\df}{\delta \phi}
\newcommand{\dkl}{\delta \kappa_{\Lambda}}
\newcommand{\lx}{\lambda}
\newcommand{\Lx}{\Lambda}
\newcommand{\ex}{\epsilon}
\newcommand{\db}{{\bar{\delta}}}
\newcommand{\lb}{{\bar{\lambda}}}
\newcommand{\lt}{\tilde{\lambda}}
\newcommand{\lr}{{\lambda}_R}
\newcommand{\lbr}{{\bar{\lambda}}_R}
\newcommand{\lk}{{\lambda}(k)}
\newcommand{\lbk}{{\bar{\lambda}}(k)}
\newcommand{\ltk}{\tilde{\lambda}(k)}
\newcommand{\mx}{{m}^2}
\newcommand{\mb}{{\bar{m}}^2}
\newcommand{\mt}{\tilde{m}^2}
\newcommand{\mr}{{m}^2_R}
\newcommand{\mk}{{m}^2(k)}
\newcommand{\Pt}{\tilde P}
\newcommand{\Mt}{\tilde M}
\newcommand{\Qt}{\tilde Q}
\newcommand{\Nt}{\tilde N}
\newcommand{\rhb}{\bar{\rho}}
\newcommand{\rht}{\tilde{\rho}}
\newcommand{\rhz}{\rho_0}
\newcommand{\yz}{y_0}
\newcommand{\rhzk}{\rho_0(k)}
\newcommand{\kx}{\kappa}
\newcommand{\kt}{\tilde{\kappa}}
\newcommand{\kk}{\kappa(k)}
\newcommand{\ktk}{\tilde{\kappa}(k)}
\newcommand{\Gammat}{\tilde{\Gamma}}
\newcommand{\Lt}{\tilde{L}}
\newcommand{\Zt}{\tilde{Z}}
\newcommand{\zt}{\tilde{z}}
\newcommand{\zh}{\hat{z}}
\newcommand{\uh}{\hat{u}}
\newcommand{\Mh}{\hat{M}}
\newcommand{\wt}{\tilde{w}}
\newcommand{\etat}{\tilde{\eta}}
\newcommand{\Gammak}{\Gamma_k}
\newcommand{\be}{\begin{equation}}
\newcommand{\ee}{\end{equation}}
\newcommand{\een}{\end{subequations}}
\newcommand{\ben}{\begin{subequations}}
\newcommand{\beq}{\begin{eqalignno}}
\newcommand{\eeq}{\end{eqalignno}}
\noindent
HD-THEP-95-27 \\
OUTP 95-27 P \\
July 1995\\
\vspace{2cm}
\begin{center}
{{ \Large  \bf
Critical equation of state from the average action
}}\\
\vspace{10mm}
J. Berges$^{\rm a}$,
N. Tetradis$^{\rm b}$ and C. Wetterich$^{\rm a}$ \\
\vspace{5mm}
a) Institut f\"ur Theoretische Physik,
Universit\"at Heidelberg \\
Philosophenweg 16, 69120 Heidelberg,
Germany\\
\vspace{3mm}
b) Theoretical Physics,
University of Oxford \\
1 Keble Rd., Oxford OX1 3NP,
U.K. \\
\vspace{3mm}
\end{center}

\setlength{\textwidth}{13cm}

\vspace{2.cm}
\begin{abstract}
{
The scaling form of the critical equation of state is computed
for $O(N)$-symmetric models. We employ a method based on an
exact flow equation for a coarse grained free energy. A suitable truncation
is solved numerically.
}
\end{abstract}
\clearpage
\setlength{\baselineskip}{15pt}
\setlength{\textwidth}{16cm}

\newpage

\setcounter{equation}{0}

A precise computation of the critical equation of state near a
second order phase transition is an old problem. From a general
renormalisation group analysis \cite{wilson} one can prove the Widom
scaling form \cite{widom}
$H = \phi^{\delta} \tilde{f}((T-T_c)/\phi^{1/\beta})$
for the relation between the magnetic field $H$, the magnetisation
$\phi$ and the difference from the critical temperature $T-T_c$.
In several models the critical exponents $\beta$ and $\delta$ have been
computed with high accuracy \cite{zinn} but the scaling function
$\tilde{f}$ is more difficult to access.
Previous attempts include an
expansion in $4-\epsilon$ dimensions in second
order in $\epsilon$ (third order for the Ising model)
\cite{zinn}. A particular difficulty for a direct computation in three
dimensions arises from the existence of massless Goldstone modes
in the phase with spontaneous symmetry breaking for models with continuous
symmetry (e.g.\ Heisenberg models with $O(N)$ symmetry for $N>1$).
They introduce severe infrared problems within perturbative or
loop expansions.

Recently a non-perturbative method has been proposed which can systematically
deal with infrared problems. It is based on
the average action $\Gamma_k$
\cite{average} which is a coarse grained
free energy with an infrared cutoff. More precisely
$\Gamma_k$  includes the effects
of all fluctuations with momenta $q^2 > k^2$ but not those with
$q^2 < k^2$.
In the limit
$k \rightarrow 0$ the average action
becomes the standard effective action
(the generating functional of the 1PI Green functions), while for
$k \rightarrow \infty$ it equals the classical or microscopic
action.
It is formulated in continuous space and all symmetries
of the model are preserved.
There is a simple functional integral representation \cite{average}
of $\Gamma_k$ also for $k > 0$ such that its couplings can,
in principle, also be estimated by alternative methods.

The exact non-perturbative flow equation \cite{rg}
for the scale dependence of
 $\Gamma_k$
takes the simple form of a renormalisation group improved
one-loop equation \cite{average}
\be
k \frac{\partial}{\partial k} \Gamma_k [\phi]
=\frac{1}{2} {\rm Tr} \left[\left( \Gamma_k^{(2)} [\phi] + R_k \right)^{-1}
 k \frac{\partial}{\partial k} R_k\right].
\label{one} \ee
The trace involves a momentum integration and summation over
internal indices. Most importantly, the relevant infrared
properties appear directly in the form of the exact inverse
average propagator $\Gamma_k^{(2)}$, which is the matrix of
second functional derivatives with respect to the fields.
There is always only one momentum integration - multi-loops
are not needed - which is, for a suitable cutoff function
$R_k(q^2)$
(with $R_k(0) \sim k^2$, $R_k(q^2 \to \infty) \sim e^{-q^2/k^2}$),
both infrared and ultraviolet finite.

The flow equation (\ref{one}) is a functional differential
equation and an approximate solution requires a truncation.
Our truncation is the lowest order in a systematic derivative
expansion of $\Gamma_k$ \cite{average,expon,morris}
\be
\Gamma_k = \int d^dx \left\{
U_k(\rho) + \frac{1}{2} Z_k \partial^{\mu} \phi_a
\partial_{\mu} \phi^a \right\}.
\label{two} \ee
Here $\phi^a$ denotes the $N$-component real scalar field and
$\rho = \frac{1}{2} \phi^a \phi_a$. We keep for the
potential term the most general $O(N)$-symmetric
form $U_k(\rho)$ since $U_0(\rho)$ encodes the equation of state.
The wavefunction renormalisation
is approximated by one $k$-dependent parameter $Z_k$.
Next order in the derivative expansion would be the
generalization to a $\rho$-dependent wavefunction
renormalisation $Z_k(\rho)$ plus a function
$Y_k(\rho)$ accounting for a possible different
index structure of the kinetic term for $N \geq 2$
\cite{average,expon}.
Going further would require the consideration of
terms with four derivatives and so on.
Concerning the equation of state for the present model,
the omission of higher derivative terms in the average action
typically generates an uncertainty of the order of the anomalous
dimension $\eta$.
The main reason is that for $\eta=0$ the kinetic term in the
$k$-dependent inverse propagator must be exactly proportional to
$q^2$ both for $q^2 \rightarrow 0$ and
$q^2 \rightarrow \infty$.
For the three-dimensional scalar theory
$\eta$ is known to be small and the derivative expansion
is, therefore, expected to give a reliable approximation.
This holds for arbitrary constant
``background'' field $\phi^a$. Similar, although less
stringent, arguments indicate a weak
$\rho$-dependence of the kinetic term. For the
scaling solution for $N=1$ this weak $\rho$-dependence has
been established explicitly \cite{morris}.

In this letter we compute the effective potential (Helmholtz free energy)
$\lim_{k \to 0}U_k(\rho)\equiv U(\rho)$
for the $O(N)$-model
directly in three dimensions from a solution of eqs.\ (\ref{one}),
(\ref{two}).
We extract the Widom scaling form
of the equation of state and give semi-analytical expressions
for $N=1$ and $N=3$.
Its asymptotic behavior yields the universal critical
exponents and amplitude ratios.
An alternative parametrisation of the equation
of state in terms of renormalised quantities is used in order
to compute universal couplings.

For a study of the behavior in the vicinity of the phase transition
it is convenient to work
with dimensionless renormalised fields
\be
\rht = Z_k k^{-1} \rho,\qquad u_k(\rht) = k^{-3} U_k(\rho(\rht)).
\label{three} \ee
With the truncation of eq.\ (\ref{two}) the exact
evolution equation for $u'_k \equiv \partial u_k/\partial \rht$
\cite{average,expon}
reduces to the partial differential equation
\beq
\frac{\partial u'_k}{\partial t} =~&\bigl(-2 + \eta\bigr) u'_k
+\bigl(1 + \eta\bigr)\rht u_k''
\nonumber \\
&- \frac{(N-1)}{4\pi^2} u_k'' l^3_1\bigl(u_k';\eta\bigr)
- \frac{1}{4 \pi^2}\bigl(3 u_k'' + 2 \rht u_k'''\bigr)
l^3_1\bigl(u_k'+2 \rht u_k'';\eta\bigr),
\label{four}
\eeq
where $t = \ln \left( k/\Lx \right)$, with $\Lx$
the ultraviolet cutoff of the theory.
The anomalous dimension $\eta$ is given in our truncation by
\cite{average,expon}
\be
\eta = - \frac{\partial}{\partial t} \ln Z_k
=\frac{2}{3 \pi^2} \kx \lx^{2} m^3_{2,2}(2 \lx \kx).
\label{five} \ee
with $\kx$ the location of the minimum of the potential, $u'_k(\kx)= 0$,
and $\lx$ the quartic coupling,
$u''_k(\kx)= \lx$.
The ``threshold'' functions $l^3_1$ and $m^3_{2,2}$
result from the momentum integration on the
r.h.s.\ of eq.\ (\ref{one}) and
account for the decoupling
of modes with effective mass larger than $k$.
They equal constants of order one for vanishing arguments and decay fast
for arguments much larger than one.
For the choice
of the cutoff function $R_k$ employed here their explicit form
can be found in refs. \cite{expon,num}.

To obtain the equation of state one has to solve
the partial differential equation (\ref{four}) for $k \to 0$.
Algorithms adapted to the numerical solution of eq. (\ref{four}) have
been developed previously \cite{num} and we refer to this work
for details.
The integration starts at some short
distance scale
$k^{-1}=\Lx^{-1}$ ($t=0$) where the average potential is equal to
the microscopic or classical potential (no integration of fluctuations
has been performed). We start with a quartic classical potential
parametrized as
$u'_{\Lx}(\rht) = \lx_\Lx (\rht - \kx_\Lx)$.
In the phase with spontaneous symmetry breaking the order
parameter $\rho_0=\lim_{k \to 0} Z_k^{-1} k \kappa$ takes a
non-vanishing value. In the symmetric phase the order parameter
vanishes, i.e.\ $\rho_0=0$ for $k = 0$. The two phases are
separated by a scaling solution for which $u_k'(\rht)$ becomes
independent of $k$.
For any given $\lx_\Lx$ there is a critical value $\kx_{cr}$
for which the evolution leads to the scaling solution.
A measure of the distance from the
phase transition is the difference
$\dkl = \kx_{\Lambda} - \kx_{cr}$.
If $\kx_{\Lambda}$ is interpreted as a function of temperature,
the deviation $\dkl$
is proportional to the deviation from the critical temperature, i.e.\
$\dkl = A(T) (T_c- T)$ with $A(T_c) > 0$.

The external field $H$ is related to the derivative
of the effective potential $U'= \partial U/\partial \rho$ by
$H_a = U' \phi_a$.
The critical equation of state relating
the temperature, the external field and the
order parameter can then be written in the scaling
form ($\phi=\sqrt{2 \rho}$)
\be
\frac{U'}{\phi^{\delta-1}} =  f(x),
\qquad x=\frac{-\dkl} {\phi^{1/\beta}}
\label{six}
\ee
with critical exponents $\delta$ and $\beta$.
For $\phi \to \infty$
our numerical solution for $U'$ obeys $U' \sim \phi^{\delta-1}$ to
high accuracy. The inferred value of $\delta$ is displayed in the
table, and we have checked the scaling relation
$\delta=(5-\eta)/(1+\eta)$.
The value of the critical exponent $\eta$ is obtained from eq.
(\ref{five}) for the scaling solution \cite{expon}.
We have also verified explicitly that
$f$ depends only
on the scaling variable $x$ for the value of $\beta$ given in the table.
In figs.\ \ref{sym} and \ref{brok} we plot log$(f)$
and log$(df/dx)$
as a function of log$|x|$ for $N=1$ and $N=3$.
Fig.\ \ref{sym} corresponds to the symmetric phase $(x > 0)$ and fig.\
\ref{brok} to the phase with spontaneous symmetry breaking
$(x < 0)$.

One can easily extract the asymptotic behavior from the logarithmic
plots.
The curves become constant both for
$x \to 0^+$ and $x \to 0^-$ with the same value, consistently with the
regularity of $f(x)$ at $x=0$. For the universal function one
obtains
\be
\lim\limits_{x \to 0} f(x) = D  \label{hd}
\ee
and $H = D \phi^{\delta}$ on the critical isotherm.
For $x \to \infty$ one observes
that log$(f)$
becomes a linear function of log$(x)$ with constant slope $\gamma$.
In this limit the universal function takes the form
\be
\lim\limits_{x \to \infty} f(x) = (C^+)^{-1} x^{\gamma} , \label{uc}
\ee
or $\lim_{\phi \to 0}U'=
(C^+)^{-1}|\dkl|^{\gamma}\phi^{\delta-1-\gamma/\beta}=\bar{m}^2$,
and we have verified the
scaling relation
$\gamma/\beta=\delta-1$ .
One observes that the zero-field magnetic susceptibility,
or equivalently the inverse unrenormalised squared mass
$\bar{m}^{-2}=\chi$, is
non-analytic for $\dkl \to 0$ in the symmetric phase:
$\chi = C^+ |\dkl|^{-\gamma}$. In this phase the correlation length
$\xi = (Z_0 \chi)^{1/2}$, which is
equal to the inverse of the renormalised mass $m_R$,
behaves as $\xi=\xi^+|\dkl|^{-\nu}$
with $\nu=\gamma/(2-\eta)$.

In the phase with spontaneous symmetry breaking $(x < 0)$
the plot of log$(f)$
fig.\ \ref{brok} shows a singularity for $x=-B^{-1/\beta}$, i.e.
\be
f(x=-B^{-1/\beta})=0. \label{seven}
\ee
The order parameter for $H=0$ therefore behaves as
$\phi = B (\dkl)^{\beta}$.
Below the critical temperature
the longitudinal and transversal
magnetic susceptibilities $\chi_L$ and $\chi_T$ are different for $N > 1$
$(f'=df/dx)$
\be
\chi_L^{-1} = \frac{\partial^2 U}{\partial \phi^2}
= \phi^{\delta-1} \bigl(\delta f(x) - \frac{x}{\beta}f'(x)\bigr),
\qquad
\chi_T^{-1} = \frac{1}{\phi}\frac{\partial U}{\partial \phi}
= \phi^{\delta-1} f(x). \label{eight}
\ee
This is related to the existence of massless Goldstone modes in the
$(N-1)$ transverse directions which imply that
the transversal susceptibility diverges for vanishing external field.
Fluctuations of
these massless modes also induce a divergence of the zero-field
longitudinal susceptibility. This can be seen from the singularity of the plot
of log$(f')$ for $N=3$
in fig.\ \ref{brok}. The first derivative of the
universal function with respect to $x$ vanishes as $H \to 0$, i.e.\
$f'(x=-B^{-1/\beta})=0$ for $N \ge 2$.
For $N=1$ there is a non-vanishing constant value for
$f'(x=-B^{-1/\beta})$ with a finite zero-field susceptibility
$\chi = C^- (\dkl)^{-\gamma}$ where
$(C^-)^{-1}=B^{\delta-1-1/\beta}f'(-B^{-1/\beta})/\beta$.
For a non-vanishing physical infrared cutoff
$k$ the longitudinal susceptibility remains finite also
for $N \ge 2$: $\chi_L \sim (k\rho_0)^{-1/2}$.
In the ordered phase
the correlation length for $N=1$ behaves as
$\xi=\xi^-(\dkl)^{-\nu}$ and, also for $N > 1$,
the renormalised minimum $\rho_{0R}= Z_0 \rho_0$ of the potential $U$
scales as
$\rho_{0R}=E (\dkl)^{\nu}$.

The amplitudes of singularities near the phase transition $D$, $C^{\pm}$,
$\xi^{\pm}$, $B$ and $E$ are shown in the table. They are not
universal since different short distance physics will result in
different wavefunction renormalisations $Z_{\varphi}$ and
$Z_{\varphi^2}$. All models in the same universality class can,
however, be related by a multiplicative rescaling of $\phi$ and
$\dkl$ (or $T_c-T$) resulting in $x \to c_x x$ and
$f \to c_f f$.
Ratios of amplitudes which are invariant
under this rescaling are universal. We display the universal
combinations $C^+/ C^-$, $\xi^+/\xi^-$,
$R_{\chi}=C^+ D B^{\delta-1}$,
$\tilde{R}_{\xi}=(\xi^+)^{\beta/\nu} D^{1/(\delta+1)} B$
and $\xi^+ E$  in the table.

The asymptotic behavior observed for the universal function can be
used to obtain a semi-analytical expression for $f(x)$.
We find the following
fit to reproduce the numerical values for both
$f$ and $df/dx$ within 1\% deviation
(apart from
the immediate vicinity of the zero of $f$
for $N=3$, cf.\ eq.\ (\ref{sixt})):
\be
f_{fit}(x)=D \bigl(1+B^{1/\beta} x \bigr)^a
\bigl(1+\Theta x \bigr)^{\Delta}
\bigl(1+c x \bigr)^{\gamma-a-\Delta} , \label{nine}
\ee
with $c=(C^+ D B^{a/\beta} \Theta^{\Delta} )^{-1/(\gamma-a-\Delta)}$.
The parameter $a$ is determined by the order of the pole of $f^{-1}$ at
$x=-B^{-1/\beta}$, i.e.\ $a=1$ $(a=2)$ for $N=1$ $(N>1)$. The fitting
parameters are chosen as $\Theta = 0.569 $ ($1.312$) and $\Delta = 0.180$
$(-0.595)$ for $N=1$ $(3)$.

There is an alternative parametrisation of the equation of state in terms
of renormalised quantities. In the symmetric phase ($\dkl < 0$)
we consider the dimensionless quantity
\be
F(s) = \frac{U_R'}{m_R^2} = C^+ x^{-\gamma} f(x), \qquad
s=\frac{\rho_R}{m_R} = \frac{1}{2}(\xi^+)^3(C^+)^{-1} x^{-2\beta}
\label{ten}
\ee
with
$\rho_R = Z_0 \rho$ and $U_R^{(n)} = Z_0^{-n} U^{(n)}$.
The derivatives of $F$ at $s=0$ yield the
universal couplings
\be
\frac{d F}{d s}(0) = \frac{U_{R}''(0)}{m_R} \equiv \frac{\lambda_R}{m_R},
\qquad
\frac{d^2 F}{d s^2}(0) = U_{R}'''(0) \equiv \nu_R \label{elev}
\ee
and similarly for higher derivatives.
They determine the behavior of $f$ for $x \gg 1/2$
\be
f(x)= (C^+)^{-1} x^{\gamma} + \frac{1}{2} \frac{\lambda_R}{m_R}
(\xi^+)^3 (C^+)^{-2} x^{\gamma-2\beta} + \frac{1}{8} \nu_R
(\xi^+)^6 (C^+)^{-3} x^{\gamma-4\beta} + \ldots  \label{thirdt}
\ee
In the ordered
phase $(\dkl > 0)$ we consider the ratio
\be
G(\tilde{s}) = \frac{U_{R}'}{\rho_{0R}^2} = \frac{1}{2} B^2 E^{-3}
(-x)^{-\gamma} f(x),
\qquad
\tilde{s} = \frac{\rho_R}{\rho_{0R}} = B^{-2} (-x)^{-2\beta}.\label{fourt}
\ee
The values for the universal couplings
\be
\frac{d G}{d \tilde{s}}(1) = \frac{U_{R}''(\rho_{0R})}{\rho_{0R}}\equiv
\frac{\hat{\lambda}_R}{\rho_{0R}},
\qquad
\frac{d^2 G}{d \tilde{s}^2}(1)=U_{R}'''(\rho_{0R})\equiv
\hat{\nu}_R \label{fivet}
\ee
as well as $\lambda_R/m_R$ and $\nu_R$
are given in the table.
One observes that for $N>1$ the renormalised quartic coupling
$\hat{\lambda}_R$ vanishes in the ordered phase.
This results from the presence of massless fluctuations.
For $x$ near $-B^{-1/\beta}$ the scaling function is approximated by
\be
f(x) = E^3 B^{-6} (-x)^{\gamma} \bigl((-x)^{-2\beta}-B^2 \bigr)
\bigl( 2 B^2 \frac{\hat{\lambda}_R}{\rho_{0R}}
+ \hat{\nu}_R \bigl((-x)^{-2\beta}-B^2 \bigr) \bigr)
+ \ldots \label{sixt}
\ee

In summary, our numerical solution of eq.\ (\ref{four}) gives a very
detailed picture of the critical equation of state. The numerical
uncertainties are estimated by
comparison of results obtained through two independent integration
algorithms \cite{num}.
They are small, typically less than $0.3 \%$ for critical exponents
and  $1 - 3 \%$ for amplitudes. The scaling relations between the
critical exponents are fulfilled within a deviation of $2 \times 10^{-4}$.
The dominant quantitative error stems from the
truncation of the exact flow equation and is related to the size of
the anomalous dimension $\eta \simeq 4 \%$.
This is consistent with the fact that
the critical exponents and amplitudes calculated here
typically deviate by a few percent from the more precise values
obtained by other methods \cite{zinn}. If the equation of state is
needed with a higher accuracy one has to extend the
truncation beyond the level of the present work.

\newpage

\section*{Tables}

\begin{table} [h]
\renewcommand{\arraystretch}{1.5}
\hspace*{\fill}
\begin{tabular}{|c|c|c|c|c|c||c|c|c|c|c|c|}     \hline

$N$
& $\beta$
& $\gamma$
& $\delta$
& $\nu$
& $\eta$
& $\lr / m_R$
& $\nu_R$
& $\hat{\lambda}_R / \rho_{0R}$
& $\hat{\nu}_R$
&
\\ \hline
1
& 0.336
& 1.258
& 4.75
& 0.643
& 0.044
& 9.69
& 108
& 61.6
& 107
&
\\ \hline
3
& 0.388
& 1.465
& 4.78
& 0.747
& 0.038
& 7.45
& 57.4
& 0
& $\simeq$ 250
&
\\ \hline \hline
$$
& $C^+$
& $D$
& $B$
& $\xi^+$
& $E$
& $C^+/C^-$
& $\xi^+/\xi^-$
& $R_{\chi}$
& $\tilde{R}_{\xi}$
& $\xi^+ E$
\\ \hline
1
& 0.0742
& 15.88
& 1.087
& 0.257
& 0.652
& 4.29
& 1.86
& 1.61
& 0.865
& 0.168
\\ \hline
3
& 0.0743
& 8.02
& 1.180
& 0.263
& 0.746
& -
& -
& 1.11
& 0.845
& 0.196
\\ \hline
\end{tabular}
\hspace*{\fill}
\renewcommand{\arraystretch}{1}
\caption[]%
{Parameters for the equation of state.}
\end{table}

\section*{Figures}

\renewcommand{\labelenumi}{Fig. \arabic{enumi}}
\begin{enumerate}
\item  
: Logarithmic plot of $f$ and $df/dx$ for $x > 0$.
\label{sym}
\item  
: Logarithmic plot of $f$ and $df/dx$ for $x < 0$.
\label{brok}
\end{enumerate}

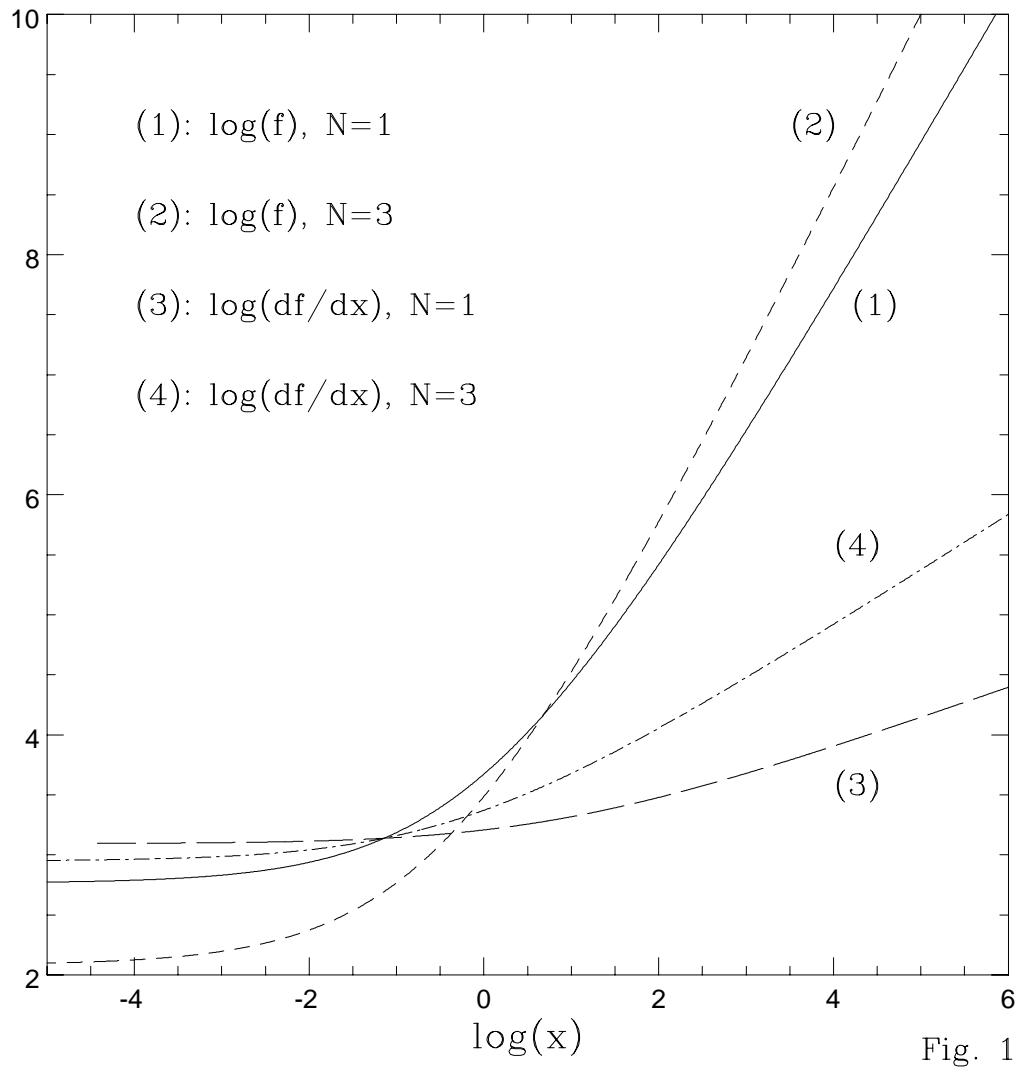
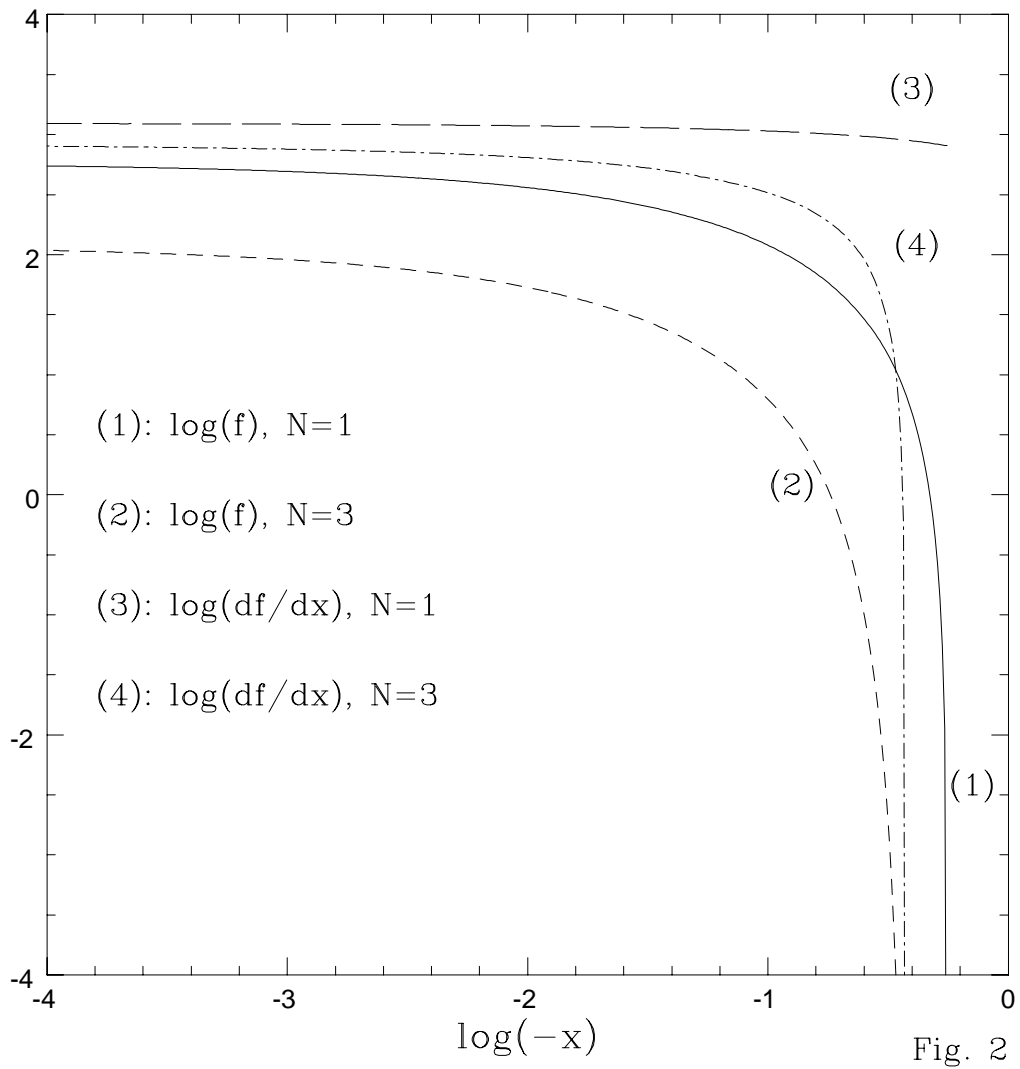
\end{document}